\documentclass[
showpacs,
 amsmath,amssymb,longbibliography,aps,prb,
 superscriptaddress,
 10pt,twocolumn
]{revtex4-2}

\usepackage[usenames,dvipsnames]{color}
\usepackage{graphicx,textcase} 
\usepackage{dcolumn,overpic} 
\usepackage{bm,color}
\usepackage[T1]{fontenc}
\usepackage{ae,aecompl}
\usepackage{chngcntr,gensymb}
\usepackage[para]{threeparttable}
\usepackage{etoolbox}
\usepackage{lipsum}
\usepackage[bottom]{footmisc}
\usepackage{setspace}

\AtEndEnvironment{table*}{\vskip10pt}{}{}

\usepackage{url}
\usepackage{hyperref}
\hypersetup{colorlinks=true,breaklinks,linkcolor=blue,urlcolor=blue,citecolor=blue}

\begin{document}

\title{Dynamical correlations in single-layer CrI$_3$}

\author{Yaroslav~O.~Kvashnin}
\address{\mbox{Division of Materials Theory, Department of Physics and Astronomy, Uppsala University, Box 516, SE-751 20 Uppsala, Sweden}}

\author{Alexander N. Rudenko}
\email{a.rudenko@science.ru.nl}
\address{\mbox{Institute for Molecules and Materials, Radboud University, Heyendaalseweg 135, 6525 AJ Nijmegen, The Netherlands}}

\author{Patrik Thunstr{\"o}m}
\address{\mbox{Division of Materials Theory, Department of Physics and Astronomy, Uppsala University, Box 516, SE-751 20 Uppsala, Sweden}}

\author{Malte R\"osner} 
\address{\mbox{Institute for Molecules and Materials, Radboud University, Heyendaalseweg 135, 6525 AJ Nijmegen, The Netherlands}}

\author{Mikhail I. Katsnelson}
\address{\mbox{Institute for Molecules and Materials, Radboud University, Heyendaalseweg 135, 6525 AJ Nijmegen, The Netherlands}}

\begin{abstract}
Chromium triiodide is an intrinsically magnetic van der Waals material down to the single-layer limit.
Here, we provide a first-principles description of finite-temperature magnetic and spectral properties of monolayer (ML) CrI$_3$ based on fully charge self-consistent density functional theory (DFT) combined with dynamical mean-field theory, revealing a formation of local moments on Cr from strong local Coulomb interactions. 
We show that the presence of local dynamical correlations leads to a modification of
the electronic structure of ferromagnetically ordered CrI$_3$. In contrast to conventional DFT+$U$ calculations, we find that the top of the valence band in ML CrI$_3$ demonstrates essentially different orbital character for minority and majority spin states, which is closer to the standard DFT results.
This leads to a strong spin polarization of the optical conductivity upon hole doping, which could be verified experimentally.
\end{abstract}

\date{\today}
\maketitle

\section{Introduction}

Conventional electronic devices are at the limit of their possible miniaturization, and  alternatives have to be sought.
Magnetism and magnetic materials can provide the key to make the leap to drastically improve memory storage and to speed up logical operations.
Spintronics \cite{spintronics-review}, which uses the elementary magnetic moments (spins) in magnetic materials to store information, is already a well-established technology, which is vastly based on ferromagnetic materials.
Thus, if one could scale the latter down to the size of atomically thin layers, an extremely high information density could be achieved.

In general, magnetism is, however, hardly compatible with the two-dimensional (2D) character of a material.
The Mermin-Wagner theorem \cite{PhysRevLett.17.1133} states that for spins coupled via short-ranged isotropic Heisenberg interactions, enhanced thermal fluctuations in 2D make the magnetic order unstable at any temperature. Yet, as long as there is a gap in the spin excitations, induced by some kind of anisotropy in the system, the magnetic order can be stabilized (see Ref.~\cite{IKK} and references therein). In the case of infinitely strong anisotropy corresponding to the 2D Ising model, where the spins are assumed to be one-dimensional objects, an exact solution is known \cite{PhysRev.65.117} hosting a long-range order below some critical temperature. Unfortunately, the limit of strong anisotropy and Ising-like spins is obviously unrealistic for typical magnetic materials, and in the case of weak anisotropy the magnetic ordering temperature is furthermore suppressed by thermal fluctuations in comparison with exchange parameters. This suppression is, however, just logarithmic, which in principle allows again for sizable ordering temperatures \cite{IKK}.  
Even though the recipe for stabilizing 2D magnetism was known before, there were until recently no 2D magnets in which the magnetism had an intrinsic origin and was not induced by point defects (see, e.g., Refs.~\onlinecite{Cervenka2009, Nair2012, Gonzalez-Herrero437}).

In 2017 it was demonstrated that a few layer and even a single layer exfoliated from weakly coupled bulk magnetic van der Waals (vdW) materials preserved the magnetic order \cite{Gong2017, Huang2017}.
In Ref.~\cite{Huang2017}, chromium iodide (CrI$_3$) was shown to exhibit magnetic order in its monolayer (ML) form up to temperatures of 45 K, which is similar to the corresponding bulk value of 61 K.
Each layer of CrI$_3$ consists of a Cr atom honeycomb network, surrounded by edge-sharing I octahedra. 
The local Cr magnetic moments have an out-of-plane orientation, and their values are very close to nominal $S=\frac{3}{2}$, which is expected for Cr$^{3+}$ ions.

Another peculiar detail of CrI$_3$ regards its thickness dependence: 
It was shown that although monolayers, trilayers, and bulk samples are ferromagnetic, the bilayers have an antiferromagnetic (AFM) order \cite{Huang2017}. 
This interlayer AFM coupling also manifests itself when a few layers of CrI$_3$ are sandwiched between, e.g., graphite contacts \cite{Song1214}. Moreover, it has been shown that the magnetic order of a bilayer system can be influenced not only by an external magnetic field, but also by an electric field \cite{cri3-efield-switch,jiang_controlling_2018}. The nature of the interlayer coupling is currently not well understood.
For a general introduction to the field, we refer the interested reader to recent reviews \cite{kostya,soriano_rev}.

CrI$_3$ is a semiconductor with an electronic band gap of 1.2 eV, which is in the lower range of the 1--3-eV band gaps of the chromium trihalides \cite{Dillon-optics-CrX3, abramchuk2018}. 
This family of materials are useful for magneto-optics and magnetic tunnel junctions, where they are already known to perform extremely well \cite{photolum-cri3,doi:10.1126/sciadv.abg8094,Cheng2021}.
Thus it is of great importance to understand the microscopic origins behind both the optical and magnetic properties of these materials. 


In general, first-principles calculations play an important role in the field of emerging 2D magnets.
They not only are used to model the already existing materials, but also have shown potential for predicting new promising systems \cite{PhysRevB.79.115409, Ma-ACSNano,Mounet2018,PhysRevB.102.024441}.
There have been a number of mainly density-functional-theory-(+$U$)-based \cite{lsdau} first-principles electronic structure studies of the 0-K ground-state properties of bulk and few-layer  CrI$_3$ \cite{C5TC02840J,Lado_2017, PhysRevB.98.144411, PhysRevB.100.205409, olsen_2019, PhysRevB.99.104432},
including the stacking dependence of the interlayer exchange coupling \cite{sivadas2018,SORIANO2019113662}, the suggestion of pathways to stabilize skyrmions \cite{skyrmions-CrI3}, and electric-field switching of the magnetization \cite{Su_rez_Morell_2019}.
Conventional DFT predicts that monolayer CrI$_3$ has a band gap of about 0.9 eV \cite{PhysRevB.98.144411} which is quite close to the experimental optical gap of 1.2 eV for bulk \cite{Dillon-optics-CrX3}. This suggests that the gap is likely defined by quasiparticle states. However, quasiparticle self-consistent $GW$ calculations for both bulk and ML CrI$_3$ predict instead much larger values, around 2.0--3.0 eV depending on the computational scheme and the presence or absence of spin-orbit coupling \cite{PhysRevB.101.241409,PhysRevB.104.155109}. 
Recent $GW$ calculations combined with Bethe-Salpeter calculations propose an alternative description, suggesting that there are in-gap states in the infrared range associated with strongly bound excitons \cite{gw-bse-cri3,D0TC01322F,acharya2021excitons}. 
These clear differences demonstrate that the electronic correlations are essentially strong in this material and that they have to be addressed in detail.

Here, we address the problem of finite-temperature magnetism and electronic excitations in single-layer CrI$_3$ from first principles.
For this purpose, we employ density functional theory (DFT) combined with dynamical mean-field theory (DMFT) \cite{infdim-DMFT}.
Within this approach, we calculate electronic spectra of the ferromagnetically ordered CrI$_3$ and analyze in detail its ground-state optical and transport properties.

The application of the DFT+DMFT approach \cite{Anis1997,LK1998,Kot_RMP2} to 2D systems is limited to a few studies. The authors of Ref.~\cite{PhysRevLett.123.236401} studied pressure-induced metal-insulator transitions in layered $M$P$X_3$ ($M$=$\{$Mn,Ni$\}$), focusing on the properties of the paramagnetic phase.
They report quite different behaviors of these seemingly similar materials and find a number of additional interesting phenomena such as structural and spin-state transitions induced by compression. A DMFT study of ML VSe$_2$ \cite{Kim_2020}
demonstrated that local Coulomb correlations provide an important contribution to the formation of ferromagnetism in this material, which is not reproduced for pristine VSe$_2$ by standard DFT calculations \cite{Memarzadeh_2021}.
Reference \cite{PhysRevB.103.035137} reports that local electronic correlations leads to the enhancement of orbital magnetization in layered ferromagnet VI$_3$. 
The bulk phase of CrI$_3$ 
was recently studied using DFT+DMFT with the multiorbital iterated perturbation theory impurity solver \cite{PhysRevB.102.195130}.
The perturbative solver produced qualitatively different spectral functions [density of states (DOS)] when starting from a ferromagnetic (FM) or nonmagnetic DFT band dispersion. In the former case the correlation effects were strongly suppressed, yielding a DFT-like DOS, while in the latter case the system became a narrow-gap Mott insulator.
Whether these observations are intrinsic to the FM and paramagnetic bulk phase of CrI$_3$ or due to the perturbative nature of the applied impurity solver is at this point an open question.

\section{Computational details}

The lattice parameters as well as atomic positions of ML CrI$_3$ were entirely relaxed using the projector-augmented-plane-wave-based code Vienna \emph{ab initio} simulation package {\sc vasp} \cite{paw,vasp} employing the Perdew-Burke-Ernzerhof exchange-correlation (xc) functional revised for solids (PBESol) \cite{PhysRevLett.100.136406}, which often gives better structural properties compared with the original PBE \cite{gga-pbe} functional.
To avoid any interactions between adjacent layers due to 3D periodic boundaries, we have added a 20-\AA-thick vacuum.
The plane-wave energy cutoff was set to 450 eV, and the $k$-point grid contained 17$\times$17$\times$1 points. More details about the crystal structure of ML CrI$_3$ obtained with PBESol can be found in Ref.~\cite{PhysRevB.102.115162}.
Since we are focused on single-layer CrI$_3$ in this paper, we do not use any van der Waals corrections. 
Once the structure was optimized, the correlated electronic structure was computed using the full-potential linear muffin-tin orbital (LMTO)-based code {\sc rspt} \cite{rspt-web,rspt-book} employing the PBE flavor of generalized gradient approximation.

The localized Cr-$3d$ orbitals were constructed via projections onto muffin-tin spheres (so-called ``MT heads'') by the DFT bands within an energy window of $[-10, 3.4 ]$ eV around the Fermi level.

To describe the local Coulomb interaction between the Cr $d$ electrons, we utilize DFT+$U$ as well as DFT+DMFT calculations.
In the DFT+$U$ calculations we set the Hubbard $U$ and Hund's $J$ parameters to 2.92 and 0.9 eV, respectively. 
The former is close to the $U$ estimated from linear response calculations ($U$=2.65 eV) \cite{C5CP04835D}. 
The value of Hund's $J$ is a typical choice for $3d$ systems that was taken from constrained DFT~\cite{ldau1}.
This estimate corresponds to a fully screened Coulomb repulsion, which also includes screening effects from the correlated Cr-$d$ states themselves, since this is not accounted for in the DFT+$U$ calculations.
These self-screening effects are, however, explicitly taken into account in our DMFT impurity solver, so for the DFT+DMFT calculations we use the Coulomb interaction parameters estimated from constrained random phase approximation (cRPA) calculations \cite{PhysRevB.70.195104}.
In the cRPA approach we calculate $\hat{U} = \hat{v} (1 - \hat{v} \hat{\Pi}_r)^{-1}$ from the bare interaction $\hat{v}$ and the rest polarization $\hat{\Pi}_r = \hat{\Pi}_f - \hat{\Pi}_d$, which includes all screening processes (described by $\hat{\Pi}_f$) except those taking place within the correlated subspace (described by $\hat{\Pi}_d$).
$\hat{\Pi}_f$ and $\hat{\Pi}_d$ are here calculated from spin-independent DFT calculations, and $\hat{\Pi}_d$ is defined by restricting all virtual transitions to bands with mostly Cr-$d$ character.
To this end, we use a recent cRPA implementation by Kaltak within {\sc vasp} \cite{KaltakcRPA}.
The resulting $\hat{U}$ is finally projected to Cr-$d$ states from a $d+p$ Wannier basis. The latter is constructed by projecting the DFT Kohn-Sham states to Cr-$d$ and I-$p$ orbitals using the {\sc wannier90} package \cite{Wannier90} and without performing a maximal localization. 
This results in a Coulomb tensor $U_{\alpha\beta\gamma\delta}$ in the Wannier basis of the Cr-$d$ states. 
We finally define $U = 1/N_U \sum_\alpha U_{\alpha\alpha\alpha\alpha} = 4.1\,$eV and $J = 1/N_J \sum_{\alpha \neq \beta} U_{\alpha\beta\alpha\beta} = 0.58\,$eV as the orbital averages of the intraorbital density-density and interorbital Hund's exchange matrix elements, respectively. Here, $N_U=5$ and $N_J=20$ are the weighting factors corresponding to the number of relevant matrix elements in the sum.  
Further details can be found in Ref.~\cite{soriano_environmental_2021}.

The combination of DFT and DMFT introduces a double-counting error in the single-particle Hamiltonian: The Cr-$d$ Coulomb interaction is explicitly added in DMFT although the DFT Hamiltonian already implicitly includes this interaction to some extent.
The choice of the corresponding double-counting (DC) correction can be quite important for the spectral properties \cite{Karolak-NiO-dc}.
Here, we use the atomic limit \cite{dc-fll, FLL-DC, lsdau} yielding the following shift of the noninteracting bands: 
\begin{eqnarray}
\mu_{\mathrm{DC}} = \frac{E_{\mathrm{DC}}}{\partial n_{\sigma}}= U(n-\frac12)-J(\frac{n}{2}-\frac12),
\end{eqnarray}
where $n$ is the total occupation of the projected correlated orbitals.
In principle, $n$ can either be found self-consistently or be fixed to a nominal value. The latter choice stabilizes the self-consistent cycle. 
Nominally, Cr$^{3+}$ is supposed to have $n=3$ electrons in its $d$ shell, but DFT calculations suggest that this occupation is closer to 4 \cite{Kashin_2020}. 
This indicates a rather strong hybridization of the Cr-$d$ states which we capture here by choosing $n=3.5$, yielding a chemical potential shift of $\mu_{\mathrm{DC}}$=7.635 eV, which was fixed throughout the calculation.

The ferromagnetic state within DMFT was stabilized by starting from a nonmagnetic DFT solution and breaking the symmetry between two spin channels, as was done in, e.g., Ref.~\cite{Fe-Ni-magQMC}.
If the system has a tendency to order magnetically and the temperature is sufficiently low, then the spin splitting stays finite during the DMFT self-consistency cycle.
In the case of single-layer CrI$_3$ the experimental ordering temperature is rather low (45 K).
Since it is very challenging to apply a quantum Monte Carlo DMFT solver to a multiorbital system at 45 K, we have opted to use an exact diagonalization (ED) solver, whose details can be found in Ref.~\cite{patrik-TMO}.
The temperature in our simulations was set to 32 K to ensure that we fall into the magnetically ordered phase, and the
DFT+DMFT calculations were performed in a fully charge self-consistent manner \cite{csc-dmft,rspt-csc}.

It is also worth mentioning that we have considered the entire $d$ shell when solving the impurity problem.
In principle, Cr$^{3+}$ in CrI$_3$ has a close to octahedral environment, and the nominal occupation is $N=3$, which makes it tempting to consider just the subset of effective $t_{2g}$ orbitals.
However, studies of interatomic Heisenberg exchange interactions clearly suggest that the effective $e_g$ electrons provide the necessary ferromagnetic superexchange interaction channels, which overcome the antiferromagnetic $t_{2g}$-derived contribution \cite{PhysRevB.99.104432, Kashin_2020}.
Thus it is impossible to self-consistently obtain the correct ferromagnetic order without taking into account these nominally empty states.

\section{Results and Discussion}

\subsection{Spectral properties}
\begin{figure}[t]
\includegraphics[width=0.95\columnwidth]{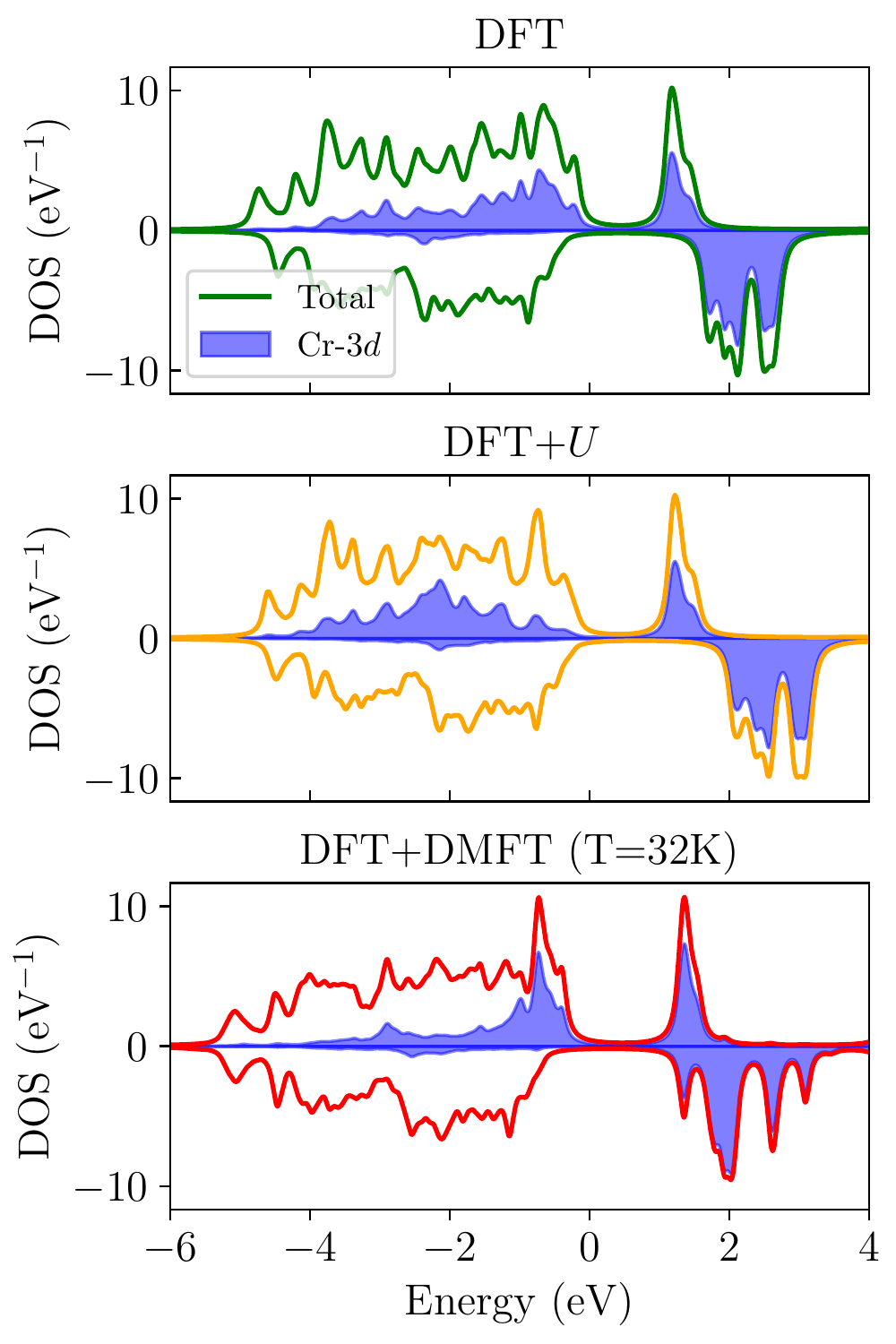} 
\caption{Comparison between the spectral densities obtained using different methods for ferromagnetic ML CrI$_3$. The Fermi level ($E_F$) is at zero energy.}
\label{compare-doses}
\end{figure}

Note that the DFT+$U$ calculations presented here were performed starting from a nonmagnetic DFT (PBE) in order to make the setup as similar to the DFT+DMFT calculations.
It is worth noting that it is more common to perform (s)DFT+$U$ calculations, starting from spin-polarized electron density \cite{lsdau}.
For instance, this approach was adopted in a previous study on ML CrI$_3$ \cite{Kashin_2020}.
The difference between DFT+$U$ and (s)DFT+$U$ is primarily related to the nature of the DC correction. In the former case, the DC potential is the same for both spin channels, because one starts from a nonpolarized density, whereas in the latter case it is different for spin up and spin down.
The resulting spin-dependent part of the DC is supposed to compensate for the exchange splitting provided by the xc functional.
Recently, there have been a number of studies showing that the total energies obtained by the two approaches can be substantially different \cite{lda-lsda, PhysRevB.93.205110, PhysRevB.97.184404}.

The low-temperature DFT+DMFT calculation predicts a stable ferromagnetic order in CrI$_3$ ML.
The resulting many-body ground state has about $n$=4.2 electrons, in good agreement with plain DFT \cite{Kashin_2020}.
It is characterized by the following expectation values of the angular momenta operators: $J= 3.46$, $L=2.82$, and $S=1.72$.
From the pure ionic picture, Cr$^{3+}$ is expected to have a spin moment of $S=\frac32$.
However, due to strong hybridization effects, we can only approximately discuss these states in terms of atomic multiplets. 
The analysis of the low-lying many-body states reveals four levels of $S\approx\frac32$ origin, which are split due to the internal magnetic field of the material.
The energy splitting between these levels is just 7 meV, which means that very low temperatures or an applied magnetic field are needed to fully saturate the magnetization.
Even at $T=32$ K, the lowest state ($S_z=\frac32$) has a weight of 0.97, whereas the remainder 0.03 comes from the first excited state with $S_z=\frac12$.

It is interesting to analyze the difference between the results obtained from DFT, DFT+$U$, and DFT+DMFT calculations.
All computational schemes predict a total magnetic moment of 6 $\mu_B$ per unit cell, containing two Cr atoms.
In DFT, DFT+$U$, and DFT+DMFT the $3d$ states of Cr carry a moment of 2.96, 3.4, and 3.07 $\mu_B$, respectively.
\begin{figure}[t]
\includegraphics[width=1.0\columnwidth]{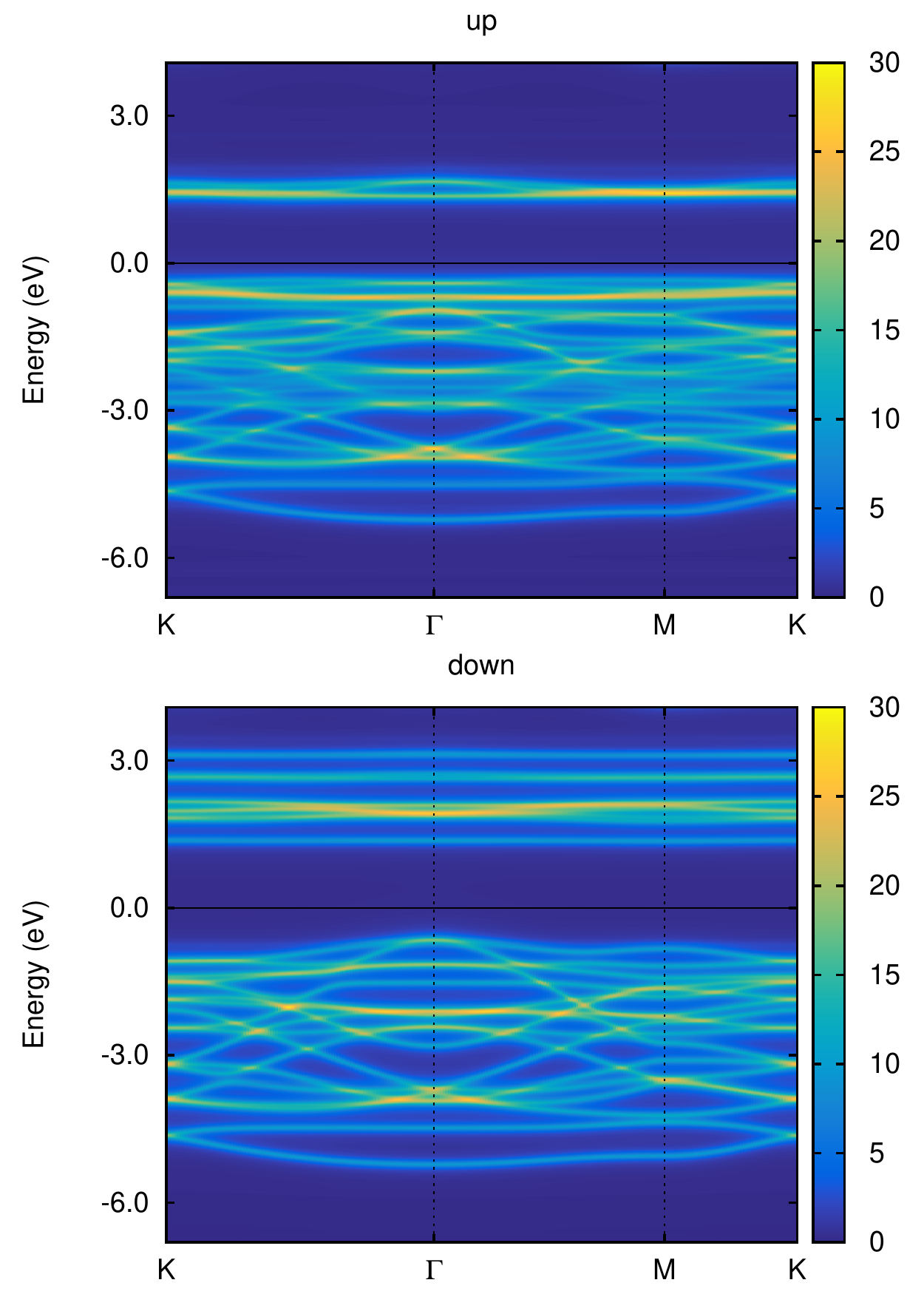} 
\caption{Spin-up and spin-down parts of the charge self-consistent DFT+DMFT-derived spectral functions for ML CrI$_3$.}
\label{dmft-spec}
\end{figure}
The calculated DOSs for ferromagnetically ordered ML CrI$_3$  are shown in Fig.~\ref{compare-doses}.
First of all, one can see that the states above the Fermi level have predominantly Cr-$3d$ character.

The lowest-lying unoccupied orbitals have complete spin-up polarization in DFT and DFT+$U$ calculations. In the DFT+DMFT calculation, these states do not have a clear spin character as a result of the incomplete saturation of the many-body magnetic state. At low enough temperature, we expect the spin-down character to disappear.

While the unoccupied states are similar in all three calculations, the situation is more peculiar for the valence band. Application of static $U$ pushes the Cr-$3d$ states to lower energies with the center of gravity being close to 2 eV below $E_F$. Thus the material appears to be somewhat in between a charge-transfer insulator and a Mott insulator.

DFT+DMFT, on the other hand, contracts the density of correlated states and also places them much closer to the top of the valence band. 
The spin-up states in the vicinity of $E_F$ are primarily of Cr-$3d$ character. This is in contrast to the DFT and DFT+$U$ results, where the relative weight of I-$p$ orbitals is much more pronounced.
The spin-down states originate from I-$p$ states, which will become even more apparent when we discuss the $k$-resolved spectral function below.
Overall, DFT+DMFT makes the system look like a Mott insulator. 

There are clear differences in the obtained density of the valence states between our DFT, DFT+$U$, and DFT+DMFT calculations.
The calculated valence bands can in principle be directly compared with angular integrated soft x-ray photoemission spectra, if the cross sections of the orbitals and the background from multiple scattering events are taken into account.   
We believe that such measurements can be used to elucidate the role of local dynamical correlations in CrI$_3$ and confirm our findings.

In Fig.~\ref{dmft-spec}, we present the $k$-resolved spectral function from the magnetic DFT+DMFT.
One can see that the dispersion of the states above $E_F$ is quite low, which is typical for the well-localized states of Cr $d$ electrons. The valence band displays mostly well defined quasiparticle states.
The broadening of the spectral features is more pronounced for spin-up electrons compared with spin-down ones.
The top of the spin-down valence states has a characteristic parabolic shape around the $\Gamma$ point, which emphasizes their I-$p$ character. We therefore observe a different orbital character of spin-up and spin-down valence states near the band gap.

\subsection{Optical conductivity}

\begin{figure}[t]
\includegraphics[width=0.8\columnwidth]{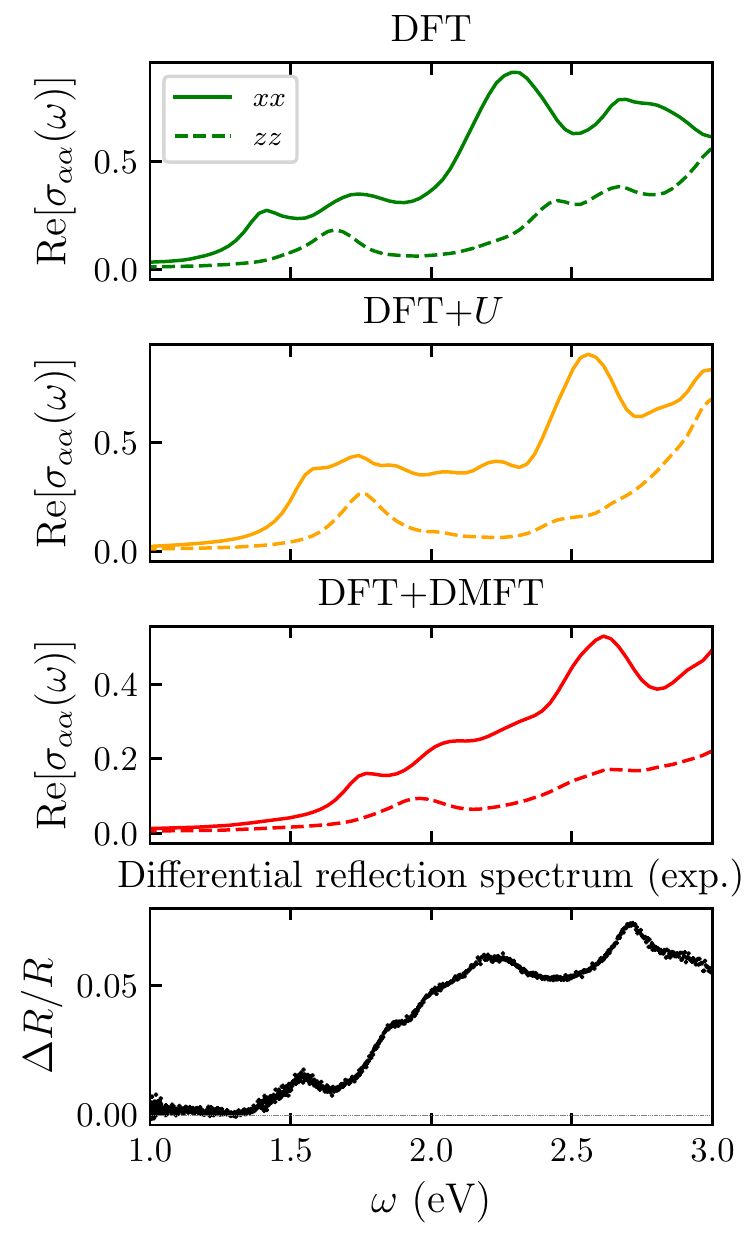} 
\caption{Comparison of the calculated optical conductivity tensor components $\sigma_{xx}$ and $\sigma_{zz}$ obtained from DFT, DFT+$U$, and DFT+DMFT. Bottom panel: experimental data from Ref.~\cite{photolum-cri3}.}
\label{fig-optics}
\end{figure}

The optical conductivity tensor $\sigma$ was calculated using the Kubo-Greenwood formalism \cite{kubo-original,Greenwood_1958,PhysRevB.9.4897,PhysRevB.74.140411}:
\begin{eqnarray}
\notag
\mathrm{Re}[\sigma_{\alpha\beta} (\omega)]= \frac{e^2}{m^2 V} \sum_{{\bf k}} \int d\omega^{\prime} \frac{f(\omega^{\prime})-f(\omega^{\prime}+\omega)}{\omega}  \\ 
\times \, \mathrm{Tr} [ p_{\alpha}({\bf k}) A({\bf k},\omega^{\prime}) p_{\beta}({\bf k}) A({\bf k},\omega^{\prime}+\omega) ],
\end{eqnarray}
where $f(\omega)$ refers to the Fermi-Dirac distribution at energy $\omega$, $\bm{p} (\textbf{k})=-ih \langle \psi_{i,\textbf{k}}|\bm{\nabla}|\psi_{j,\textbf{k}} \rangle$ are the matrix elements of the dipole (i.e., momentum) operator calculated in the LMTO basis, $V$ is the unit cell area, and $A$ is an anti-Hermitian part of the Green's function,
\begin{equation}
    A({\bf k},\omega)=\frac{i}{2}[G({\bf k},\omega)-G^\dagger({\bf k},\omega)].
\end{equation}

Note that the spectral function $A$(\textbf{k},$\omega$) of bare quasiparticles [obtained within DFT(+$U$)] has poles on the real energy axis. 
Thus we evaluate those in the complex energy plane with an imaginary offset of 0.034 eV, which results in their Lorentzian broadening.
The latter is usually chosen in order to account for various dissipation channels, such as scattering of electrons with each other or with phonons.
At the same time, the DMFT-derived spectral functions are subject to intrinsic smearing, defined by the imaginary part of the self-energy.
This means that some of the scattering channels are already taken into account and the effective imaginary offset should be different.
However, since we cannot quantify the impact of other processes, for simplicity we used the same Lorentzian broadening for all three sets of results.

The results obtained for the three computational setups are shown in Fig.~\ref{fig-optics}.
Within all three sets of results, there is a strong anisotropy in the light absorption having an in-plane or out-of-plane direction of the polarization.
The DFT-derived results are in good agreement with Ref.~\cite{C5TC02840J}.
The experimental data for bulk CrI$_3$, shown in the bottom panel, suggest the first pronounced peak to be at around 1.5 eV.
Standard DFT calculations for ML CrI$_3$ seem to reproduce this feature rather well.
However, one has to keep in mind that spin-orbit coupling, which was not taken into account here, results in a smaller band gap \cite{Lado_2017,PhysRevB.101.241409}, which is likely to shift these features down and worsen the agreement with experiment.
In this sense, the DFT+$U$ and DFT+DMFT results, which currently seem to push this peak to higher energies, might improve upon the inclusion of spin-orbit coupling. Another factor affecting the results could be nonlocal correlation effects, not considered in this paper.

\begin{figure}[t]
\includegraphics[width=0.98\columnwidth]{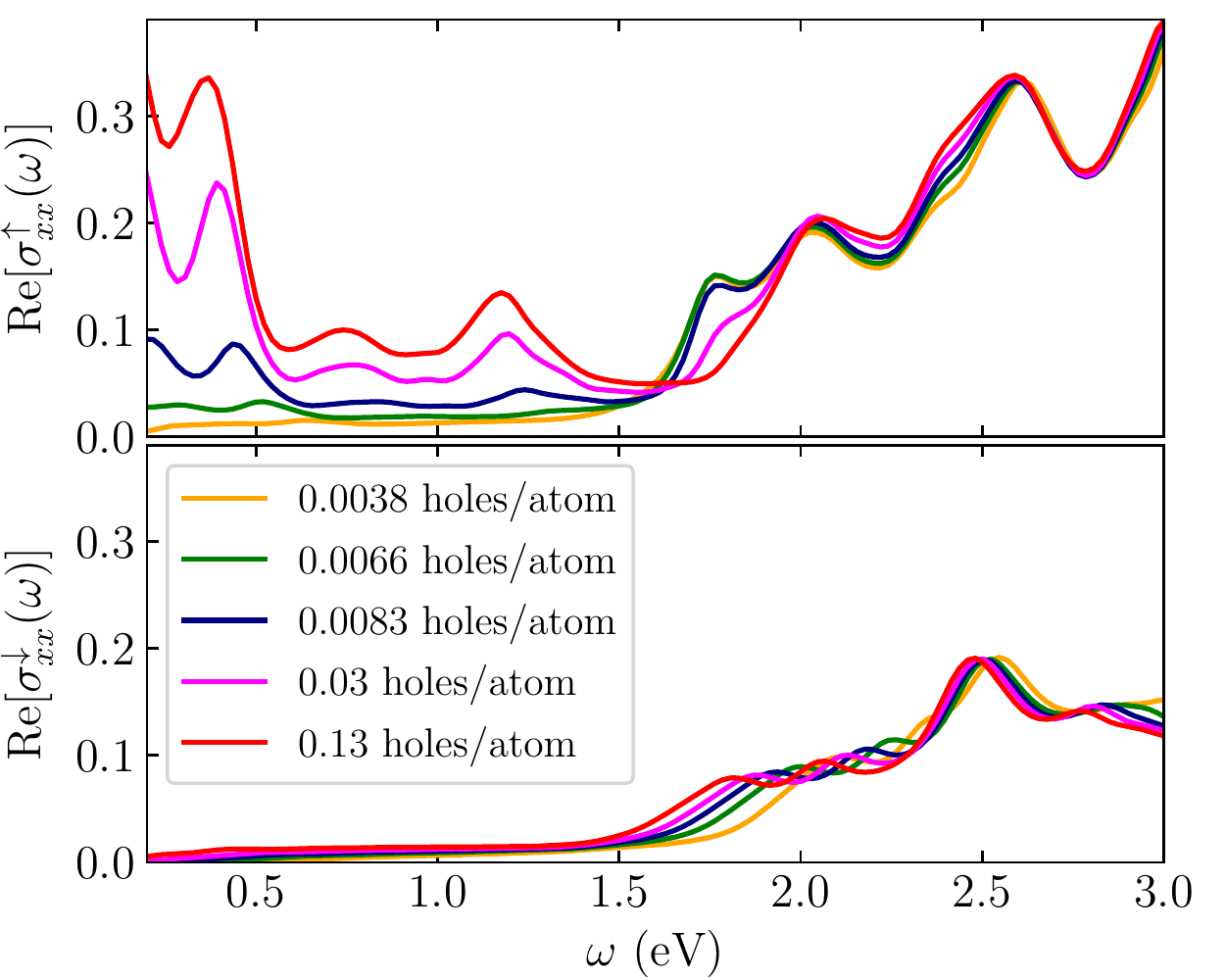} 
\caption{DFT+DMFT-derived optical conductivity $\sigma_{xx}$ projected onto spin-up (top panel) and spin-down (bottom panel) contributions, obtained for different doping concentrations.}
\label{fig-doping}
\end{figure}

Since the top of the valence band was found to be strongly spin polarized, it is natural to study the effect of hole doping on the optical absorption.
In Fig.~\ref{fig-doping}, we show the in-plane optical conductivity calculated for DFT+DMFT within the rigid band model (i.e., by shifting the chemical potential).
We see that indeed even a tiny amount of doping leads to conductivity in only one spin channel and thus effectively acts as a spin filter.
The latter idea has been suggested in prior studies based on DFT \cite{Wang_2016,YU2021147693}.
Although our calculations are performed without spin-orbit coupling being considered, earlier studies show that the valence band in Cr$X_3$ demonstrates a high degree of the spin polarization that is weakly affected by the presence of spin-orbit coupling (e.g., see Ref.~\cite{D0TC01322F}).
It is worth noting that DFT+DMFT predicts the highest degree of spin dependence of the optical absorption among the three considered computational setups, which is followed by a less pronounced effect in DFT calculations.
DFT+$U$ calculation, on the other hand, results in a comparable optical conductivity in both spin channels upon hole doping, which is in contradiction to the other two schemes.
In fact, plain DFT results appear to be more similar to those of DFT+DMFT, since the carriers are predicted to be of primarily Cr-$3d$ majority states.

\section{Outlook}

In this paper, we examined the effect of static and dynamical local correlations 
on the electronic structure of ML CrI$_3$ in the ferromagnetic phase. To this end, we used a combination of first-principles DFT-based calculations and DMFT.
We found that the presence of correlation effects primarily affects the valence band character: They shift the Cr-$3d$ band with respect to the I-$5p$ band and thus change essentially the character of the states near the band edge in comparison with mean-field approaches, such as DFT+$U$. At the same time, the description of the conduction band remains qualitatively the same. Interestingly enough, there is less difference between DFT and DFT+DMFT than between DFT and DFT+$U$. This means that for this particular compound, it is better to ignore the effects of the Hubbard $U$ than to take them into account at the static mean-field level. This situation is not unique and is also quite typical for rare-earth compounds \cite{TbN}. 

The other interesting result is a strong spin polarization to the optical conductivity upon hole doping. As follows from Fig.~\ref{fig-doping}, the main contribution results from the majority-spin states. 
Note that in Ref.~\cite{jiang_controlling_2018}, the electrostatic doping was shown to strongly affect the magnetization and the ordering temperature of few-layer CrI$_3$.
Our results suggest that this should be accompanied by a strong temperature dependence of the optical conductivity below the Curie temperature as a result of the temperature dependence of the magnetization. For the same reasons, it should be also sensitive to the value of the external magnetic field. These two results can be verified experimentally.
\\

\begin{acknowledgments}
The authors are thankful to Kyle Seyler and Xiaodong Xu for providing the experimental data from Ref.~\cite{photolum-cri3}.
Y.O.K. acknowledges the financial support from the Swedish Research Council (VR) under Project No. 2019-03569 and the G{\"o}ran Gustafsson Foundation. The work by M.I.K. and A.N.R. is supported by the European Research Council via Synergy Grant No. 854843 - FASTCORR.
The calculations were performed using the resources provided by the Swedish National Infrastructure for Computing (SNIC) at the National Supercomputing Centre (NSC).
\end{acknowledgments}

\bibliography{mybib}

\end{document}